\documentclass[12pt]{article}
\usepackage{amsmath}
\usepackage{graphicx,color}
\newcommand{\de}{\partial}
\newcommand{\be}{\begin{equation}}
\newcommand{\ba}{\begin{eqnarray}}
\newcommand{\ea}{\end{eqnarray}}
\newcommand{\ee}{\end{equation}}

\newcommand{\f}{\frac}
\newcommand{\s}{\sqrt}

\newcommand{\ti}{\tilde}
\newcommand{\ap}{\alpha}

\newcommand{\ddd}{\cdot\cdot\cdot}
\newcommand{\no}{\nonumber \\}
\newcommand{\la}{\langle}
\newcommand{\lb}{\rangle}
\newcommand{\bea}{\begin{eqnarray}}
\newcommand{\eea}{\end{eqnarray}}
\newcommand{\bes}{\begin{equation*}}
\newcommand{\beas}{\begin{eqnarray*}}
\newcommand{\eeas}{\end{eqnarray*}}
\newcommand{\bas}{\begin{array*}}
\newcommand{\eas}{\end{array*}}
\newcommand{\ees}{\end{equation*}}

\newcommand{\p}{\partial}
\newcommand{\ep}{\epsilon}

\begin{document}

\begin{titlepage}
\thispagestyle{empty}

\begin{flushright}
YITP-13-117
\\
IPMU13-0224
\\
IPM/P-2013/045
\\
\end{flushright}


\begin{center}
\noindent{\large \textbf{Holographic Geometry of cMERA
for \\ Quantum Quenches and Finite Temperature}}\\
\vspace{2cm}

Ali Mollabashi $^{a,b}$,
Masahiro Nozaki $^{b}$,
Shinsei Ryu $^{c}$
and
Tadashi Takayanagi $^{b,d}$
\vspace{1cm}

{\it
$^{a}$School of physics, Institute for Research in Fundamental Sciences (IPM), Tehran, Iran\\
$^{b}$Yukawa Institute for Theoretical Physics,
Kyoto University, Kyoto 606-8502, Japan\\
$^{c}$Department of Physics, University of Illinois at Urbana-Champaign,
1110 West Green St, Urbana IL 61801, USA\\
$^{d}$Kavli Institute for the Physics and Mathematics of the Universe,\\
University of Tokyo, Kashiwa, Chiba 277-8582, Japan\\
}

\vskip 2em
\end{center}

\begin{abstract}
We study the time evolution of cMERA (continuous MERA) under quantum quenches in free field theories. We calculate the corresponding holographic metric using the proposal in arXiv:1208.3469 and confirm that it qualitatively agrees with its gravity dual given by a half of the AdS black hole spacetime, argued by Hartman and Maldacena in arXiv:1303.1080. By doubling the cMERA for the quantum quench, we give an explicit construction of finite temperature cMERA. We also study cMERA in the presence of chemical potential and show that there is an enhancement of metric in the infrared region corresponding to the Fermi energy.
\end{abstract}

\end{titlepage}

\newpage

\section{Introduction}

Even though the AdS/CFT correspondence \cite{Maldacena}
has been confirmed in and successfully applied to many examples,
our knowledge on the basic mechanism of AdS/CFT is far from complete. Especially we need to better understand how the metric of the bulk anti de-Sitter space (AdS space) emerges from the dynamics of conformal field theories (CFTs). To expand our understandings on this fundamental question will be useful to approach a more general principle known as holography \cite{Hol}, so that we can deal with quantum gravity in spacetimes other than AdS spaces.

One interesting possibility in this direction is that the AdS/CFT may be interpreted as the real space renormalization scheme called MERA (multi-scale entanglement renormalization ansatz) \cite{MERA} as conjectured by Swingle \cite{Swingle}. This connection suggests that the spacetime in gravity can be regarded as collections of bits of quantum entanglement and explains the holographic entanglement entropy \cite{RT} in a very beautiful way. A closely related viewpoint has also been pointed out in \cite{Ra}. Moreover, in the paper
\cite{NRT}, the expression of holographic metric in the extra direction was proposed purely in terms of field theoretic data by employing a field theory limit of MERA (called cMERA i.e. continuous MERA \cite{cMERA}). See e.g.\cite{MS,BMR,MIH,Mat,SwC,EvVi,Qi} for other interesting developments in this topic. Refer also \cite{Lee} for another interesting connection between the holographic emergent metric and the renormalization group flow.

The aim of this paper is to better understand this connection between the AdS/CFT and MERA at finite temperature.  The gravity dual in AdS/CFT in this case is well-known and is described by an AdS black hole. Therefore if we understand this relation in detail, we can in principle approach still mysterious properties of black holes.  MERA at finite temperature has already been considered in \cite{Swingle,MIH,HaMa} and has argued to be described by a doubling the standard MERA for a pure state and gluing together at infrared points, which follows from the thermofield double construction. This structure nicely agrees with the geometry of external AdS black holes \cite{MaE}. Even though this description is useful to speculate the global structure of spacetime, we need to perform considerable numerical computations in order to
calculate physical quantities or entanglement structures in a specific quantum many-body systems.

Therefore, in this paper we would like to study cMERA at finite temperature. Actually, this attempt, at least at first sight, immediately faces a problem. The reason is that in the cMERA for a pure state, we first need to choose the infrared state (IR state), which has no entanglement at all between any (spatially defined) subsystems. Then we will add quantum entanglement at each length scale and in the end we will reproduce the original quantum state (UV state). However, it is not obvious at all what kind of IR pure state in the doubled Hilbert space we should choose for the cMERA at finite temperature. Since the renormalization procedure which adds the entanglement is given by a unitary transformation in cMERA, the total entropy does not change. Thus the IR state should be an entangled state and this makes its choice very ambiguous.

Nevertheless, thanks to the recent observation by Hartman and Maldacena \cite{HaMa}, a close connection between the gravity dual of quantum quench \cite{CaCaG} and that of finite temperature CFT has been pointed out.
The quantum quench is an instantaneously excited state of a given quantum system for example by suddenly changing a mass parameter \cite{CaCaG}. In cMERA we can construct such a pure state which is produced by a quantum quench in a straightforward way. The gravity dual suggests that we can construct cMERA at finite temperature by doubling the cMERA for the quantum quench and we will argue that this is indeed true by showing several evidences.

This paper is organized as follows: In section 2, we will give a brief review of (c)MERA and its holographic interpretation. In section 3, we will study quantum quenches in cMERA for free scalar field theories. In section 4, we discuss a holographic interpretation of the cMERA for quantum quenches analyzed in section 3.  In section 5, we will construct cMERA at finite temperature and discuss its properties. In section 6, we will study cMERA for free Dirac fermions. In section 7, we will compute holographic metrics for finite temperature CFTs with non-vanishing chemical potentials. In section 8, we summarize our conclusions.

\section{Brief Review of cMERA}

Here we would like to present a brief summary of the idea of MERA (multi-scale entanglement renormalization ansatz)
\cite{MERA} and its continuous formulation called cMERA \cite{cMERA}. We will also explain its holographic interpretation
following \cite{Swingle,NRT}.

\subsection{MERA}

The idea of MERA is a scheme of real space renormalization in terms of wave functions. This is in contrast with the more familiar method of Wilsonian renormalization group, where we consider the renormalization group flow in momentum space in terms of effective actions.

Suppose we want to find the ground state of a given quantum spin chain with a complicated Hamiltonian by employing a variational principle of quantum mechanics. The real space renormalization means that we coarse-grain the spin chain by combining two spins into one at each step. Let us define the (non-positive) integer $u$ which counts the steps of this coarse-graining. We describe the initial spin chain by $u=0$ and the first step of  coarse-graining is denoted by $u=-1$.
If we start with a spin chain with $N$ spins, after $n=-u>0$ steps of coarse-graining the number of spins becomes
$N\cdot 2^{u}$. In the end, it is reduced to a single spin after $\log_2 N$ steps.

We can have parameters for this coarse-graining procedure (mathematically called isometry transformation). However, even if we optimize them by minimizing the total energy, following the variational
principle, we cannot obtain a good approximation of correct ground state if the quantum spin chain does not have a mass gap. This is because in such a wave function (called tree tensor network) has much smaller amount of quantum entanglement. We can easily confirm that the entanglement entropy $S_A$ has a finite upper bound. On the other hand, we know that $S_A$ increases logarithmically with respect to the size of $A$.

\begin{figure}[ttt]
   \begin{center}
     \includegraphics[height=4cm]{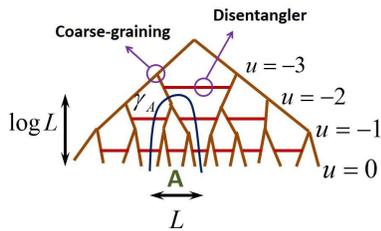}
   \end{center}
   \caption{The schematic structure of MERA.}\label{fig:MERA}
\end{figure}

To circumvent this problem, in MERA, we introduce so called disentanglers which cut bits of quantum entanglement of the original highly entangled ground state. Refer to Fig.\ref{fig:MERA}. A disentangler is a unitary transformation which acts on each of nearest neighbor spins in each coarse-graining step. If we look this procedure in an opposite way, we can start from a single spin. Then we double the number of spins and add some quantum entanglement by the unitary transformation of adjacent spins by the (dis)entanglers at each step. In the end we reproduce the correct ground state. These are the basic construction of MERA. Note also that we can generalize this formulation of $1+1$ dimensional MERA to higher dimensions in a straightforward way.

\begin{figure}[ttt]
   \begin{center}
     \includegraphics[height=4cm]{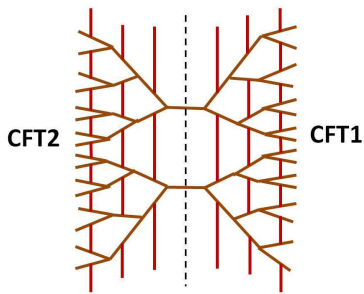}
   \end{center}
   \caption{The schematic structure of MERA at finite temperature.}\label{fig:FTMERA}
\end{figure}

  For a MERA description of a finite temperature CFT, we can remember the thermofield formalism, where the thermal state is described by a pure state in the doubled Hilbert space of the CFT. This consideration naturally leads to the MERA construction presented in Fig.\ref{fig:FTMERA} as argued in \cite{MIH,HaMa}. The entangling bonds in the middle which separates the left and right half are responsible for the entanglement between the two CFTs and thus the number of them is proportional to the thermal entropy.

\subsection{cMERA}
In order to understand field theories from the viewpoint of MERA, we need to consider a continuum limit of MERA. This is called the continuous MERA (cMERA), first presented in \cite{cMERA}. We will follow the convention of cMERA in \cite{NRT}. In cMERA, we start from the unentangled state $|\Omega\lb$ (IR state) and add the entanglement for each length scale so that we can reproduce the correct state $|\Psi\lb$ (UV state) which we want as a ground state for a given Hamiltonian. This construction is naturally understood from a continuous limit of MERA. Only apparent difference is that in cMERA the dimension of Hilbert space
(or the number of spins) does not change in each coarse-graining step. However, this can be simply understood as adding dummy states at each coarse-graining step so that the total number of spins does not change.

We define a state $|\Psi(u)\lb$ parameterized by the scale $u$. This state $|\Psi(u)\lb$
is obtained by adding the entanglement for the momentum scale $k\leq \Lambda e^{u}$ to the unentangled state $|\Omega\lb$. We choose $\Lambda=1/\ep$ to be the original UV cut off scale. If we take $u=0$, then the state includes all the entanglement and coincides with the UV state (e.g. the ground state) we are looking for i.e. $|\Psi(0)\lb=|\Psi\lb$.
On the other hand, if we set $u=-\infty$, then the state does not include any entanglement and is given by the IR state $|\Psi(-\infty)\lb=|\Omega\lb$.

If we write this procedure explicitly, we have
\be\label{unitaryTrnsf}
|\Psi(u)\lb=e^{-iuL}\cdot Pe^{-i\int^u_{-\infty}\hat{K}(s)ds}|\Omega\lb,
\ee
where $\hat{K}(s)$ denotes the process of adding the entanglement at scale
$s$ (i.e. $k=\Lambda e^{^s}$) \cite{cMERA}. The symbol $P$ means the path-ordering which
puts all operators with smaller $u$ to the right. For later convenience we also define ˜$\ti{P}$ as the one
with the opposite order. The operator $L$ is defined by the scale transformation and the factor $e^{-iuL}$ means the scale transformation at scale $u$ so that $|\Psi(u)\lb$ fits nicely with the discrete MERA description. In the language of AdS/CFT, this factor corresponds to the warp factor of the AdS metric and it is useful to redefine the state into a ``normalized'' state $|\Phi(u)\lb$ by eliminating this factor \cite{NRT} as follows:
\be
|\Psi(u)\lb=e^{-iuL}|\Phi(u)\lb.
\ee
In this formulation of cMERA,  $|\Phi(u)\lb$ is represented as the integral of disentangler action
\be
|\Phi(u)\lb=Pe^{-i\int^u_{u_{IR}}\hat{K}(s)ds}|\Omega\lb.
\ee
It is also useful to notice the relation:
\be
|\Psi(u)\lb=e^{-iuL}\ti{P} e^{-i\int^u_{0}\hat{K}(s)ds}|\Psi(0)\lb. \label{uvf}
\ee

\subsection{cMERA for Free Scalar Field Theory}

Consider the free scalar field theory in $1+1$ dimensions (with mass $m$). The time and space
coordinates are denoted by $t$ and $x$. The energy and the momentum in $x$ direction are written as $\ep$ and $k_x$, respectively. Though we can generalize most of our arguments in this paper to higher dimensions, just for simplicity we choose the two dimensional theory. We write the creation and annihilation operator of the scalar field as $a_{k_x}^{\dagger}$ and $a_{k_x}$, which satisfy $[a_{k_x},a^{\dagger}_{k'_x}]=\delta(k_{x}-k'_x)$. We define $k=|k_x|$ and then the dispersion relation is given by $\ep_k=\s{k^2+m^2}$.

In the IR limit, the ground state is described by infinitely many independent copies of　harmonic oscillators at each lattice point. The unentangled state
$|\Omega\lb$ is the ground state for harmonic oscillator Hamiltonian $H\propto \sum_{x}a^{\dagger}_xa_x$, and is defined by
$a_x|\Omega\lb=0$. In momentum space, this condition is equivalent to
\be
(\ap_k a_k+\beta_k a^{\dagger}_{-k})|\Omega\lb=0, \label{irs}
\ee
where
\ba
&&\ap_k=\f{1}{2}\left(\s{\f{M}{\ep_k}}+\s{\f{\ep_k}{M}}\right),\ \ \ \
\beta_k=\f{1}{2}\left(\s{\f{M}{\ep_k}}-\s{\f{\ep_k}{M}}\right),\ \ \
M=\s{\Lambda^2+m^2}.
\ea
To obtain (\ref{irs}), we first discretize the Hamiltonian with the
lattice constant $\ep=1/\Lambda$ in the $x$ direction. Then we simply ignore the interactions between difference lattice points and pick up only the self interactions. The ground state of this unentangled Hamiltonian is given by (\ref{irs}). Note that for the massless theory
we have $M=\Lambda$.

Assuming that the state is ``gaussian'', the disentangler $\hat{K}$ takes the following form
\be
\hat{K}(u)=\f{i}{2}\int dk_x \Gamma(ke^{-u}/\Lambda)\left(g(u)a^{\dagger}_{k_x}a^{\dagger}_{-k_x}
-g^*(u)a_{k_x}a_{-k_x}\right), \label{Ksc}
\ee
where $\Gamma(x)$ is the cut off function such that $\Gamma(x)=1$ when $x\leq 1$ and $\Gamma(x)=0$ for $x>1$. Indeed, for the ground state defined by
\be
a_{k}|\Psi(0)\lb=0,
\ee
the ansatz (\ref{Ksc}) reproduces the exact correct state if we set \cite{cMERA}
\be
g(u)=g^*(u)=\f{1}{2}\cdot \f{e^{2u}}{e^{2u}+m^2/\Lambda^2}. \label{scge}
\ee

\subsection{Excited States}

We focus on a class of excited states defined by
\be
(A_{k}a_{k_x}+B_{k} a^{\dagger}_{-k_x})|\Psi(0)\lb=0. \label{abrep}
\ee
At scale $u$, the state $|\Psi(u)\lb$ satisfies
\be
(A_k(u)a_{k_x}+B_k(u) a^{\dagger}_{-k_x})|\Psi(u)\lb=0, \label{exab}
\ee
where we assume $|A_k(u)|^2-|B_k(u)|^2=1$. It is obvious that we have
$(A_k(-\infty),B_k(-\infty))=(\ap_k,\beta_k)$ and $(A_k(0),B_k(0))=(A_k,B_k)$.

We define the $SU(1,1)$ matrix $M_k(u)$ by
\be
(A_k(u),B_k(u))=(\ap_k,\beta_k)\cdot M_{k}(u).
\ee
We can express $M_k(u)$ as
\be
M_k(u)=\left(
  \begin{array}{cc}
    p_k(u) & q_k(u) \\
    q^{*}_k(u) & p^{*}_k(u) \\
  \end{array}
\right)
\ee
with $|p_k(u)|^2-|q_k(u)|^2=1$,
where
\be
p_k(u)=\ap_k A_k(u)-\beta_k B_k^*(u),\ \ \ q_k(u)=-\beta_k A^{*}_k(u)
+\ap_k B_k(u).
\ee

%

Then we can define $2\times 2$ matrix $G(u)$ so that
\be
M_{k}(u)=\ti{P}\cdot \exp\left(-\int^u_{-\infty}ds   G(s)\Gamma(ke^{-s}/\Lambda)\right).
\ee
In particular at $u=0$ we find
\be
M_{k}(0)=\ti{P}\cdot \exp\left(-\int^0_{\log\f{k}{\Lambda}} du G(u)\right),
\ee
and this satisfies the obvious relation:
\be
M_{\Lambda}(0)=1. \label{uvlim}
\ee
For example, it is easy to see that the choice of $\hat{K}$ (\ref{Ksc}) corresponds to
\be
G(u)=\left(
\begin{array}{cc}
  0 & g(u) \\
  g^*(u) & 0
\end{array}
\right).  \label{realf}
\ee
If $g(u)$ is real valued, we can neglect the path-ordering as the group structure is abelian.
We can obtain $G(u)$ from $M_k(0)$ as follows:
\be
G(\log k/\Lambda)=k\f{dM_k(0)}{dk}\cdot M^{-1}_k(0).
\ee

\subsection{Relation to AdS/CFT}

An interesting observation is that the structure of MERA (Fig.\ref{fig:MERA}) looks very similar to
a time slice of AdS space (i.e. hyperbolic space). Indeed, it has been conjectured in \cite{Swingle} that the mechanism of AdS$_{d+2}$/CFT$_{d+1}$ is equivalent to the $d+1$ dimensional MERA. To be exact, since we need to take the continuum limit to describe the CFT, we can employ cMERA instead of MERA \cite{NRT}.

We denote the Poincare coordinate of AdS$_{d+2}$ by
\be
ds^2=\f{dz^2-dt^2+\sum_{i=1}^{d}dx_i^2}{z^2}.
\ee
Since it is known that the extra direction $z$
of the AdS space corresponds to the length scale of renormalization group flow, we can naturally
identify
\be
z=\ep e^{-u},
\ee
where $\ep=\Lambda^{-1}$ is the UV cut off (or lattice spacing) of the CFT.
More generally, a generic state in cMERA, the metric looks like
\be
ds^2=g_{uu}du^2+\Lambda^2 e^{2u}dx^2+\ddd, \label{metc}
\ee
where the omitted terms $\ddd$ involve $dt$ and we will not discuss these components below.

Now, consider the calculation of entanglement entropy $S_A$. As is explained in Fig.\ref{fig:MERA},
$S_A$ is bounded from above by the number of entangling bonds which intersect with a surface
$\gamma_A$. Here $\gamma_A$ is an arbitrary surface which surrounds the region $A$. To optimize this
bound we need to choose $\gamma_A$ which minimizes the number of bonds. This prescription looks very similar to the holographic entanglement entropy \cite{RT}, which is given by the minimal area divided by four times the Newton constant $G_N$. This observation is a very important evidence of this conjecture.

By closely studying this argument, we can relate the metric $g_{uu}$ in (\ref{metc}) to the density of disentanglers. This idea leads to the following conjectured expression of $g_{uu}$ in cMERA \cite{NRT}:
\be
g_{uu}du^2\propto 1-|\la \Phi(u)|\Phi(u+du)\lb|^2.
\ee

In particular, consider the free scalar field theory and assume the form (\ref{Ksc}).
If $g(u)$ is real, the metric component $g_{uu}$ is given by
\be
g_{uu}\propto g(u)^2. \label{AdS}
\ee
For example, for the massless theory $m=0$ we find from (\ref{scge}) that $g_{uu}$ is constant and this is consistent with the expectation that its holographic metric (\ref{metc}) coincides with a pure AdS spacetime.

\section{Quantum Quenches in cMERA and Holography}

Now we would like to study quantum quenches in cMERA for free scalar field theories and discuss a holographic interpretation.

\subsection{Quantum Quenches and Boundary States}

We would like to transform the quantum quench calculations considered by \cite{CaCaG} into the framework of cMERA. Quantum quenches are sudden excitations of a quantum system due to an instantaneous change of a Hamiltonian. For example, it is triggered by a sudden shift of a mass parameter. Since we are interested in excited states in CFTs motivated by the AdS/CFT, we consider a process where the mass parameter is changed from a non-zero value $\Delta m$ to
zero. The key idea of \cite{CaCaG} is that the excited state after such a quantum quench can be approximated by the boundary state $|B\lb$ for low energy modes. This is concretely expressed as follows:
\ba
&& |\Psi(0)\lb=e^{-\f{\beta}{4}H}|B\lb.  \label{ccstate}
\ea
The factor $e^{-\beta H/4}$ comes from the fact that for modes with energy larger than $\Delta m\sim 1/\beta$ the quantum quench has no effect and the state behaves like a vacuum. The detailed normalization of $\beta$ was chosen for a later convenience.

In the free scalar field theory we are focusing on here, it is written explicitly as
\ba
&& |\Psi(0)\lb ={\cal{N}}\cdot\exp\left(\pm\f{1}{2}\int dk_x e^{-\beta\ep_k/2}a^{\dagger}_{k_x} a^{\dagger}_{-k_x}\right)|0\lb. \label{CC}
\ea
Note that $a^{\dagger}_{k_x}$ is interpreted as the creation operator of the right-moving or left-moving mode depending on the sign of $k_x$. The signs $+$ and $-$ in front of  the integral $\f{1}{2}\int dk_x\ddd $ correspond to the boundary state for the Neumann boundary and the Dirichlet boundary condition, respectively.

This excited state (\ref{CC}) corresponds to the following choice in the class (\ref{abrep})
\be
A_{k}=\f{1}{\s{1-e^{-\beta\ep_k}}},\ \ \ B_{k}=\mp \f{e^{-\beta\ep_k/2}}{\s{1-e^{-\beta\ep_k}}}.
\ee

From now, we assume the massless case and take the dispersion relation to be $\ep_k=k$.
Then we get the function $g(u)$ in each case
\ba
&& g(u)_{N}=\f{1}{2}+\f{k\beta e^{k\beta/2}}{2(e^{k\beta}-1)},\no
&& g(u)_{D}=\f{1}{2}-\f{k\beta e^{k\beta/2}}{2(e^{k\beta}-1)}, \label{gnd}
\ea
with the identification: $k=\Lambda e^u$. Note that in both cases the function $g(u)$ are real.

As is obvious from (\ref{gnd}), we obtain the UV behavior $g(0)=1/2$ in both cases, which is the same as that of the ground state. This is simply because the excitation induced by the quantum quench has finite energy and cannot modify the UV behavior. However they have different IR behaviors. In the Dirichlet case, we find $g(-\infty)=0$, while in the Neumann case, we have $g(-\infty)=1$. It is natural that in the Dirichlet case $g(u)$ is decreased and that the IR degrees of freedom is reduced because it is similar to a large mass at $t=0$. On the other hand, our result suggests that the Neumann boundary condition increases the IR degrees of freedom. This seems to be closely related to the idea of boundary entropy \cite{AL} and the details of this connection will be an interesting future problem.

\subsection{Time-dependence in Quantum Quenches}

Since we are interested in the quantum quench triggered by the mass change, we will focus on the Dirichlet case below. As a next step, we would like to study the time-dependence.
 The time evolution of (\ref{CC}) for Dirichlet boundary condition is simply given by
\ba
&& |\Psi(0,t)\lb=e^{-\f{\beta}{4}H}|B\lb \no
&& ={\cal{N}}\cdot\exp\left(-\f{1}{2}\int dk_x e^{-\beta\ep_k/2}e^{-2i\ep_k t}a^{\dagger}_{k_x}a^{\dagger}_{-k_x}\right)|0\lb. \label{CCt}
\ea
Below we will focus on the massless case and set $\ep_k=k$ and $M=\Lambda$.

In this case, $(A_k,B_k)$ is given by
\be
A_k=\f{1}{\s{1-e^{-\beta\ep_k}}}e^{i\ep_k t+i\theta_k(t)},\ \ \ B_{k}= \f{e^{-\beta\ep_k/2}}{\s{1-e^{-\beta\ep_k}}}e^{-i\ep_k t+i\theta_k(t)},
\ee
where $\theta_k(t)$ represents the ambiguity which does not change the UV state $|\Psi(0)\lb$ defined by (\ref{exab}), though the intermediate states $|\Psi(u)\lb$ depend on $\theta_k(t)$.
Note that the identity (\ref{uvlim}) argues
\be
\theta_\Lambda (t)=-\Lambda t. \label{inic}
\ee

Now, we can choose $\theta_k(t)$ so that the diagonal parts of the matrix
$G(u)$ vanish as in (\ref{realf}). This choice is expressed as follows:
\be
\f{\de \theta_k}{\de k}=-t\coth(k\beta/2)+\f{\Lambda^2-k^2}{4(k^2+\Lambda^2)\sinh(k\beta/2)}\cdot
(4t\cos2\theta_k+\beta \sin2\theta_k). \label{constr}
\ee
We can integrate (\ref{constr}) by imposing the initial condition (\ref{inic}) and find a unique function $\theta_k(t)$. In particular at $t=0$ we simply find $\theta_k(t)=0$.

Then the component $G_{12}=g(u)$ is given by
\be
g(u)=\f{1}{2}+\f{1}{\sinh(k\beta/2)}\left(kt\sin(2\theta_k)-\f{k\beta}{4}\cos2\theta_k\right)
+O(\Lambda^{-1}), \label{qqmet}
\ee
where we expanded by assuming $k<<\Lambda$ and kept the finite term. In this limit $k<<\Lambda$,
$g(u)$ is real and so we can use the formula (\ref{AdS}).

\begin{figure}[ttt]
   \begin{center}
     \includegraphics[height=4cm]{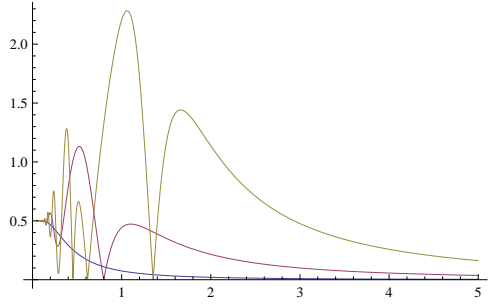}
   \end{center}
   \caption{A plot of $|g(u)|$ as a function of $z=1/k$ for $t=0$ (blue), $t=1$ (red)
   and $t=2$ (yellow).
    We chose $\beta=2$ and $\theta_0=0$ at $k=100$.}\label{fig:quenchmett}
\end{figure}

\begin{figure}[ttt]
   \begin{center}
     \includegraphics[height=4cm]{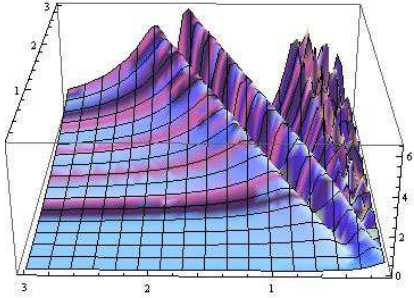}
   \end{center}
   \caption{A plot of $|g(u)|$ as a function of $z=1/k$ (horizontal coordinate) and $t$ (depth coordinate). We chose $\beta=1$ and $\theta_0=0$ at $k=100$. We plotted the region specified by $0<z<3$ and $0<t<3$.}\label{fig:quenchmet}
\end{figure}

For a large momentum $k\beta>>1$, we can easily solve (\ref{constr})
\be
\theta_k=-kt+\theta_0(t). \label{UVth}
\ee
Moreover, the boundary condition (\ref{inic}) at the UV cut off scale tells us that
$\theta_{0}(t)=0$ when $k=\Lambda$.  We plotted $|g(u)|$ in Fig.\ref{fig:quenchmett} and Fig.\ref{fig:quenchmet} in certain cases.

On the other hand, if we take the IR limit $k\to 0$, we find the solution at any $t$ behaves as:
\be
\theta_k=-A(t)k^2 +O(k^4).  \label{IRth}
\ee
The equation (\ref{constr}) allows a $O(k)$ term in the $k\to 0$ limit. However, we can numerically confirm\footnote{Moreover, we find that $\theta_k(t)$ approaches to a ladder function
when $k$ is small and $t$ is large such that $\theta_k(t)\simeq -n\pi$ for
$\f{(2n-1)\pi}{2t}<k<\f{(2n+1)\pi}{2t}$, where $n=0,1,2,\ddd$.} that this terms is vanishing under the boundary condition (\ref{inic}). This fact, combined with (\ref{constr}), leads to $A(t)=\f{\beta t}{4}$.

Even though $g(u)$ is oscillating due to the phase factor $\theta_k(t)$, this seems to be a peculiar property for free field theories. In generic interacting theories, we naturally expect $g(u)$ will be a smoother function of $u$ (or equally $k$) because different momentum modes are mixed due to interactions. Thus we would like to
replace $g(u)^2$ with its smoothed version. This clearly leads to the following behavior in the
high energy region $k\beta > O(1)$:
\be
g_{uu}=g(u)^2\simeq \f{1}{4}+ \f{a_1k^2\beta^2+a_2k^2t^2}{4\sinh^2(k\beta/2)}, \label{qmet}
\ee
where $a_1$ and $a_2$ are  certain numerical order one positive constants.

When the subsystem $A$ is the half space, we can estimate the entanglement entropy
$S_A$ at late time $t>>\beta$ by using the metric $g_{uu}$. If we subtract the entanglement entropy before the quench $t<0$, then we can estimate the increased amount $\Delta S_A$ when at late time $t>>\beta$  as follows:
\ba
\Delta S_A \sim \int^0_{-\log(\beta/\ep)}du \left(\s{g_{uu}}-\f{1}{2}\right)
+\int^{-\log(\beta/\ep)}_{-\infty} du \s{g_{uu}} \sim  \f{t}{\beta}. \label{enttt}
\ea
The metric $g_{uu}$ behaves like $\s{g_{uu}}\sim \f{kt}{\sinh(k\beta/2)}$ in the higher energy region $k\beta > O(1)$. The integration over the deep IR region i.e. $kt<<1$ does not contribute because we can confirm by using (\ref{IRth})
$\s{g_{uu}}\simeq k^2t^2$ in this region. Moreover, we can show that the integral (\ref{enttt}) for the middle energy range $O(1/t)<k<O(1/\beta)$ can be estimated again to be $O(t/\beta)$.\footnote{This can be seen as follows. In the footnote 2, we mentioned the ladder functional profile of $\theta_k$ for $k<<O(1/\beta)$. Using (\ref{constr}), we can estimate the gradient $\de_k\theta_k$ to be of order $O(\beta)$
and $O(t^2/\beta)$, for the horizontal part and vertical part of the ladder, respectively. The function $g(u)\propto \sin 2\theta_k$ has $N_p=t/\beta$ peaks in the region $O(1/t)<k<O(1/\beta)$. We can see from the mentioned ladder structure of $\theta_k$ that we have the large value $g(u)\sim O(t/\beta)$ only for the range $\Delta u=O(\beta/t)$ for each peak.
Thus we can estimate the integral (\ref{enttt}) for $O(1/t)<k<O(1/\beta)$ as
$N_p\cdot|g(u)|\cdot(\Delta u)\sim (t/\beta)(t/\beta)(\beta/t)=t/\beta$.} In this way we obtain the estimation (\ref{enttt}). This result (\ref{enttt}) reproduces the results in 2d CFT that $S_A$ is increasing linearly w.r.t $t$, as computed in \cite{CaCaG}.

Our analysis here can be generalized to higher dimensions in a straightforward way. We find $S_A$ is again a linear function of $t$ and this is consistent with the holographic result in \cite{HaMa,Liu}.
Refer also to e.g. \cite{QuenchHEE} for numerical calculations of holographic entanglement entropy under quantum quenches.

\section{Holographic Interpretation}

Now we would like to discuss a holographic interpretation. We would like to compare our cMERA construction for the quantum quench with the recent argument of the gravity dual of quantum quench \cite{HaMa}. Since we analyze the cMERA by using free field theories, our comparison will be only at qualitative level.

\begin{figure}[ttt]
   \begin{center}
     \includegraphics[height=4cm]{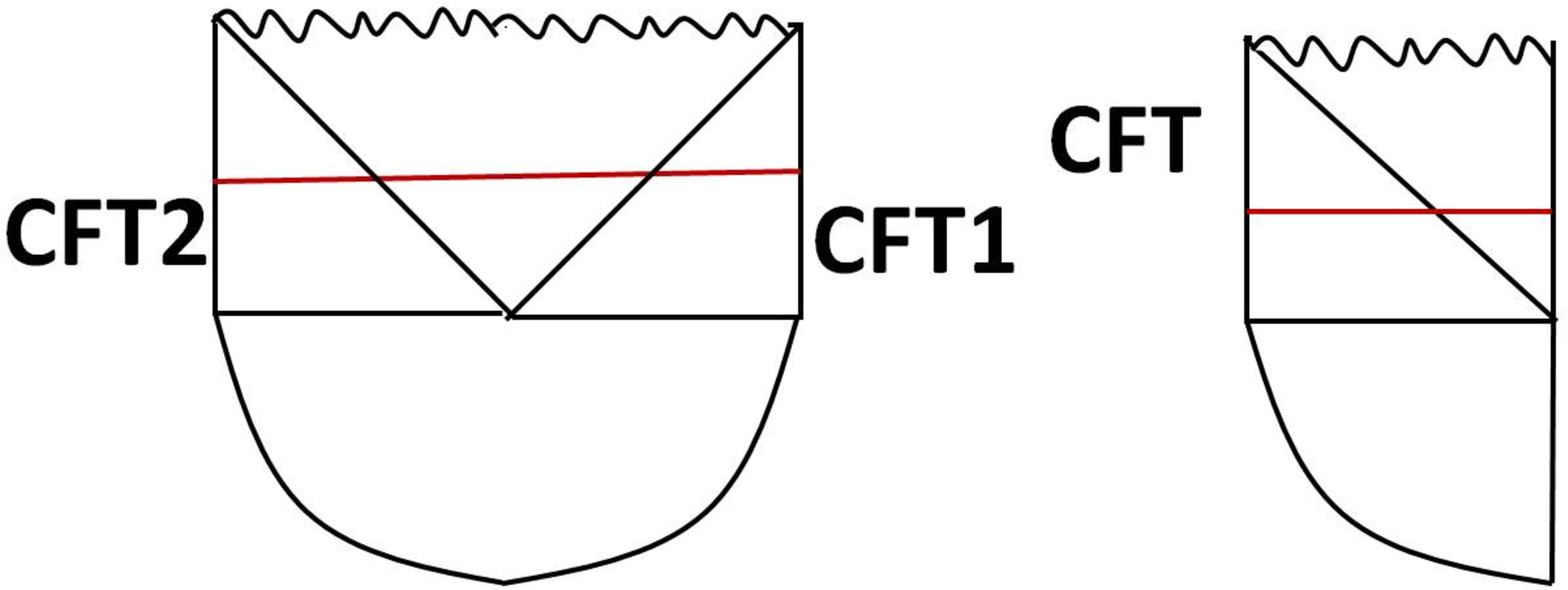}
   \end{center}
   \caption{Global Structures of AdS Schwarzschild black hole $M_{BH}$ (left) and the gravity dual of quantum quench $M_{Q}$ (right) argued in \cite{HaMa}. The red horizontal curve denotes time slices we are interested in. Following the Hartle-Hawking prescription, we treat the $t>0$ and $t<0$ region in the Lorentz and Euclidean signature, respectively.
   The diagonal lines describe the horizons of the black hole. The wavy lines in the top part represents the black hole singularities.}\label{fig:HM}
\end{figure}

Let us remember that the AdS Schwarzschild black hole, which is described by the following metric in the three dimensional case
\be
ds^2=-\f{1-z^2/z_H^2}{z^2}d\tau^2+\f{dz^2}{z^2(1-z^2/z_H^2)}+\f{dx^2}{z^2},
\ \ \left(z_H=\f{\beta}{2\pi}\right) \label{BTZ}
\ee
can be extended into a spacetime (we call this $M_{BH}$) with two boundaries. The presence of horizons separates the extended spacetime of AdS Schwarzschild black hole into four regions I, II, III and IV.  The asymptotic AdS boundaries are included in the region I and III. These two asymptotic boundaries correspond to the two CFTs: CFT$_1$ and CFT$_2$ in the thermofield description of finite temperature CFT. The regions II (future) and IV (past) are situated inside horizons. The Hartle-Hawking state of this eternal AdS black hole at time $t$ is dual to the CFT state \cite{MaE}
\be
|\Psi(t)\lb_{th} \propto\sum_{n}e^{-2itE_n}e^{-\beta E_n/2} |n\lb_{1}|n\lb_{2}, \label{thdp}
\ee
where $E_n$ and $|n\lb_{1,2}$ are the eigenvalue and eigenstate of the Hamiltonian $H_1$ and $H_2$ of the two CFTs. The dependence on the time $t$ is generated by the Hamiltonian $H_1+H_2$.
This gravity dual of this state is depicted in the left picture of Fig.\ref{fig:HM}, where the time slice is schematically written as the red curve. Notice that the time evolution with respect to $\tau$ in (\ref{BTZ}) corresponds to the Hamiltonian $H_1-H_2$, which does not change
$|\Psi(0)\lb_{th}$.

It was argued in \cite{HaMa} that the holographic dual of the quantum quench state (\ref{ccstate}) is dual to a half of extended spacetime of the AdS Schwarzschild black hole (we call this $M_{Q}$) as depicted in the right picture of Fig.\ref{fig:HM}. This is realized by introducing a real time-like boundary in the region II and IV. This identification can be thought of as one example of the AdS/BCFT correspondence \cite{AdSBCFT}. The time evolution of this quantum quench state is well described by using the time $t$ instead of $\tau$ and is dual to the black hole creation at $t=0$. At late time, the region inside the horizon at time $t$ expands so that its size is proportional to $t$.

Let us turn to the spacetime obtained in cMERA. First remember that we made the particular choice of $\theta_k$ (\ref{constr}) by requiring the diagonal parts of
the matrix $G(u)$ are vanishing. This is originally due to a technical reason
that we want to calculate the holographic metric by using the simple formula (\ref{AdS}). As proposed in \cite{NRT}, the choice of $\theta_k$ corresponds to that of the time slice on which we define $|\Psi(u)\lb$. We argue that our choice of $\theta_k$ (\ref{constr}) corresponds to one of generic time slices like the one depicted in the red curve of the left picture of Fig.\ref{fig:HM}.

By using the coordinate $z=\ep e^{u}$, we find from our previous analysis that at $t=0$ (just after the quench), the metric given by (\ref{AdS}) behaves like
\ba
g_{uu}(=z^2g_{zz})&\simeq &\mbox{a positive const.} \ \ \ (0<z<<\beta),\no
&\simeq &\mbox{0}\ \ \ (z>>\beta).  \label{tzerog}
\ea
The reason why we find $g_{uu}$ vanishes for large $z$ is because we started with a mass gapped theory for $t<0$. In the dual gravity geometry $M_Q$, this corresponds to the fact that the spacetime ends at $z=\beta$.

One may think this a little strange because the black hole usually leads to a large extensive contribution to the holographic entanglement entropy, while the cMERA result (\ref{tzerog}) shows that the entanglement entropy will be reduced compared with the CFT ground state. However, this is not any contradiction since the holographic entanglement entropy $S_A$ for a large enough interval $A$ (width$>>\beta$) is given by the area of disconnected planes which simply extend from the AdS boundary to the black hole horizon:
\be
S_A\propto \int^{\beta}_{\ep}\f{dz}{z\s{1-z^2/\beta^2}}=\log \f{\beta}{\ep}.
\ee
Thus this is smaller than that for the CFT (corresponding to the limit $\beta\to \infty$) and thus the result looks like a confining geometry, suggesting the mass gap, though we actually consider the AdS BH solution. This is possible because the geometry $M_Q$ is defined with the new boundary where $\gamma_A$ can simply end on \cite{AdSBCFT}. In this way, this holographic behavior agrees with our $g(u)$ for the quantum quench.

Next, let us consider the time evolution. At late time $t>>\beta$ the geometry in cMERA is divided into three regions:
\ba
&& \mbox{(i)}\  0<z<<\beta:  g_{uu} \simeq \mbox{a positive const.}, \no
&& \mbox{(ii)}\  O(\beta)<z<O(t): g_{uu}\propto t,\no
&& \mbox{(iii)}\  z>>t:  g_{uu}\simeq 0.
\ea
The region (i) obviously corresponds to the asymptotically AdS region.
 The region (ii), which is responsible to the entanglement entropy (\ref{enttt}), nicely corresponds to the inside horizon region in the gravity dual $M_Q$. Both grows linearly under time evolution. The region (iii) can be negligible because the metric is very small.

We can see from Fig.\ref{fig:quenchmet} that the excitations are approximately included in the light cone $z<t$. This suggests that these propagations can be related to the gravitational waves as they should be if we assume the equivalence between cMERA and AdS/CFT. Let us study this behavior more closely.
The centers of peaks $z=z_c(t)$ of $g(u)$ grow linearly under time-evolution $z_c\simeq vt$ with some velocity $v$. This can be easily seen from Fig.\ref{fig:quenchmet} and can also be explained from the behavior $g(u)\propto \sin\theta_k\sim \sin(2kt)$ derived from (\ref{qqmet}) and (\ref{UVth}) in the large $t$ limit. If we regard each of peaks as massive objects
(strings), then this time evolution is interpreted that they are falling toward the IR region
$z\to \infty$. This is consistent with the AdS space, where a massive object falls into the horizon due to the gravitational force. In this way, at a qualitative level, the time evolution of excited states in cMERA can be understood from the Newton force in the gravity dual. However, note that the velocity $v$ of this falling is not always one in our cMERA, though it is easy to see\footnote{This can be shown from (\ref{constr}). If we take the average of $\theta_k$ (denoted by $\bar{\theta}_k$ with respect to the time $t$, we find at late time
$\de\bar{\theta}_k/\de k\simeq -t\coth(k\beta/2)$. This can be solved with the boundary condition (\ref{inic}) as $\bar{\theta}_k=-\f{2t}{\beta}\log[2\sinh(k\beta/2)]$. From this we find
the peak points $\bar{\theta}_k=\pi/2,3\pi/2,\ddd$ satisfies $z_c\simeq vt$ with $v<\f{4}{\pi}$.}
 that $v$ is bounded from the above as $v<\f{4}{\pi}$, which might be analogous to the light cone.

\section{Finite Temperature cMERA}

Now we would like to move on to the construction of cMERA at finite temperature.
 As we already explained, there is an interesting connection to the quantum quench suggested by the gravity dual as sketched in Fig.\ref{fig:HM}. Even though the construction of the IR state is not obvious from the beginning for a finite temperature cMERA, we will employ this useful fact to find the correct IR state as we will show below.

\subsection{cMERA for free scalars at finite temperature}

For the free scalar at finite temperature $T=\beta^{-1}$, the pure state (\ref{thdp}) in the thermofield description (i.e. in the doubled Hilbert space) at time $t$ is written as
\ba
|\Psi(0,t)\lb_{th}
&=&{\cal N}\cdot e^{-it(H_1+H_2)}\cdot \prod_{k}\sum_{n_k=0}^\infty e^{-\beta \ep_k n_k/2}
|n_k\lb_1 |n_k\lb_2
\no
& =&{\cal N}\cdot \exp\left(\int dk e^{-\f{\beta \ep_k}{2}}e^{-2i\ep_k t}a^{\dagger}_k
\ti{a}^{\dagger}_k  \right)|0\lb |\ti{0}\lb.  \label{thwave}
\ea
Here $\ti{a}_k$ is the creation operator of scalar field in the thermofield double.

 An important observation is that (\ref{thwave}) is reduced to (\ref{CC}) for the Dirichlet boundary condition by the projection
\be
\ti{a}_k \to a_{-k},\ \ \  |0\lb |\ti{0}\lb \to |0\lb.
\ee

This relation between the quantum quench and the finite temperature CFT is precisely matches with that in their gravity duals. In this way, we find that we can choose the disentangler $\hat{K}(u)$ for the UV state (\ref{thwave}) precisely in the same way as that in the quantum quench case (\ref{CCt}):
\be
|\Psi(0,t=0)\lb_{th}=P e^{-i\int^0_{u_{IR}}\hat{K}(s)ds}\otimes P e^{-i\int^0_{u_{IR}}\hat{\ti{K}}(\ti{s})d\ti{s}}
|\Omega(\beta)\lb.
\ee
Here $|\Omega(\beta)\lb$ is a state in CFT$_1\otimes$CFT$_2$ and is highly entangled between these two CFTs. The entanglement entropy between these two CFTs for the pure state $|\Omega(\beta)\lb$ in the doubled Hilbert space is obviously equal to the thermal entropy of the free scalar in the single Hilbert space because the disentangler action is a unitary transformation in each of the two CFTs.
Note that $g(u)$ in $\hat{K}(u)$ is exactly the same as that for the quantum quench i.e.
the equations like (\ref{constr}) and (\ref{qqmet}) remain the same.
Since we know the UV state $|\Psi(0,t=0) \rangle$, which is given by (\ref{thwave}), the above relation uniquely determines the IR state $|\Omega(\beta)\lb$.

\subsection{Metric at Finite Temperature}

Now we would like to analyze the metric $g_{uu}$. In order to be consistent with the gravity dual result, this metric for the finite temperature CFT should be equal to that of the quantum quench.
To see this, we first need to find a relation between $g_{uu}$ and $g(u)$ in our finite temperature case.

First we use the description in terms of a single Hilbert space by tracing out the other one and consider its (mixed state) density matrix. We denote the density matrix at scale $u$ by $\rho_u$. We can define $\ti{\rho}_u$ in a similar way as (\ref{uvf}):
\be
\ti{\rho}_{u}=e^{iuL}\rho_u e^{-iuL}=
\ti{P}e^{-i\int^u_{0}\hat{K}(s)ds}\rho_{0}P e^{i\int^u_{0}\hat{K}(s)ds}.
\ee
In our current setup, the UV density matrix reads
\be
\rho_0=\mbox{Tr}_{CFT_2}|\Psi(0,t)\lb_{th}\la\Psi(0,t)|=
\f{1}{Z(\beta)}e^{-\beta\int dk_x \ep_k a^{\dagger}_{k_x}a_{k_x}}, \label{denz}
\ee
where $Z(\beta)=\prod_{k_x}(1-e^{-\beta\ep_k})^{-1}$ is the standard partition function of the scalar field at finite temperature.

In this density matrix formalism, a natural definition of the metric $g_{uu}$ is as follows (see \cite{NRT})
\be
g_{uu}du^2=\f{1}{2{\cal N}_{m}}\mbox{Tr}(\rho_{u+du}-\rho_u)^2,
\ee
where ${\cal N}_{m}$ is the normalization factor. For pure states in $d+1$ dimensional free scalar field theories, this is given by
\be
{\cal N}_{m}=c_m\cdot \int_{k<\Lambda e^u} d^d k, \label{ewe}
\ee
as found in \cite{NRT}. Note that ${\cal N}_m$ is proportional to the effective phase volume.
The coefficient $c_m$ is independent from $u$.
We find
\ba
\mbox{Tr}(\rho_{u+du}-\rho_u)^2=-\mbox{Tr}\left([\hat{K}(u),\rho_u][\hat{K}(u),\rho_u]\right)du^2.
\ea

Now we consider the disentangler given by the form (\ref{Ksc}) and assume that $g(u)$ is real.
Then we obtain
\ba
&& g_{uu}= -
\f{1}{2{\cal N}_{m}}\mbox{Tr}\left([\hat{K}(u),\rho_0][\hat{K}(u),\rho_0]\right) \no
&& =\f{g(u)^2}{8{\cal N}_{m}Z(\beta)^2}\int_{k,p\leq \Lambda e^u} d^d k d^d p\
(e^{2\beta(\ep_p-\ep_k)}+1)(1-e^{-2\beta\ep_p})(1-e^{-2\beta\ep_k})
\mbox{Tr}[a_p a_{-p} a^{\dagger}_k a^{\dagger}_{-k}(\rho_0)^2] \no
&& =c_m\f{Z(2\beta)}{Z(\beta)^2}\cdot g(u)^2,
\ea
where we employed
\be
\mbox{Tr}[a_p a_{-p} a^{\dagger}_k a^{\dagger}_{-k}(\rho_0)^2]=\left(\delta^d(k-p)+\delta^d(k+p)\right)\cdot \f{2}{(1-e^{-2\beta \ep_k})^3}.
\ee

In this way we find $g_{uu}\propto g(u)^2$ even for the finite temperature CFT.\footnote{
In the appendix A, we will present another definition of metric, which also leads to the result
$g_{uu}=g(u)^2$.} Even though the (UV) density matrix is time-independent like (\ref{denz}), the structure of disentangler is time-dependent precisely as that for the quantum quench.
Therefore we can calculate the corresponding time-dependent metric $g_{uu}(u,t)$ as $g(u,t)^2$ and the result is just the same as that for the quantum quench, which was computed in (\ref{qmet}). This is consistent with the gravity dual \cite{HaMa} where the metric for the quantum quench and the finite temperature CFT are the same.

\section{cMERA for free Fermions under Quantum Quenches}
In this section we will consider the quantum quench for a free fermion field theory. As we have explained in previous section that we can double the Hilbert space of cMERA to construct the finite temperature cMERA. Since this trick can be equally applied to the free fermion discussed here, we will not mention the details of
finite temperature cMERA in this section. For free fermions, cMERA has been worked out for the zero temperature ground state in version 1 of \cite{cMERA} and \cite{NRT}. For simplicity we will consider a Dirac fermion in a 1+1 dimensional space as
\bea
S_F=\int dtdx\left[i\bar{\psi}\left(\gamma^t\p_t+\gamma^x\p_x\right)\psi-m\bar{\psi}\psi\right],
\eea
where the $\gamma$ matrices are chosen to be $\gamma^t=\sigma_3$ and $\gamma^x=i\sigma_2$ in terms of Pauli matrices. Also we use the standard definition for $\bar{\psi}=\psi^\dagger\gamma^t$. The Hamiltonian of this theory after performing the Fourier transformation is given by
\bea\label{HamiltonianF}
H=\int dk\begin{bmatrix}\psi^\dagger_1(k)&\psi^\dagger_2(k)\end{bmatrix}\begin{bmatrix}m&k\\k&-m\end{bmatrix}\begin{bmatrix}\psi_1(k)\\ \psi_2(k)\end{bmatrix}.
\eea
Canonical quantization leads  to the following anti-commutation relations
\bea\label{quantization}
\{\psi_1(k),\psi^\dagger_1(p)\}=\{\psi_2(k),\psi^\dagger_2(p)\}=\delta(k-p).
\eea
In the following we will first define cMERA for free fermions and continue with applying it to a quantum quench between the zero and finite temperature cases.

\subsection{cMERA for Free Fermion}
We will simply follow the definition of IR state $|\Omega\lb$ in \cite{cMERA}:
\bea
\psi_1(k)\left| \Omega\right \rangle=\psi^\dagger_2(k)\left| \Omega\right \rangle=0.
\eea
As we will come back later, we need some modification of $|\Omega\lb$ for the UV region to get sensible results and thus we will focus only on the IR physics which can be studied from $|\Omega\lb$ defined in the above.

The true physical ground state of the Hamiltonian \eqref{HamiltonianF} is
\bea\label{Fgrnd}
\chi_1(k)\left|0\right \rangle=\chi^\dagger_2(k)\left|0\right \rangle=0.
\eea
The $\chi_1(k)$ and $\chi_2(k)$ fields are the eigenvectors of \eqref{HamiltonianF}
\bea\label{eigenvector}
\chi_1(k)=\alpha_k\psi_1(k)-\beta_k\psi_2(k),\hspace{1cm}\chi_2(k)=\beta_k \psi_1(k)+\alpha_k\psi_2(k),
\eea
where
\bea
\alpha_k=\frac{-k}{\sqrt{k^2+\left(\epsilon_k-m\right)^2}},
\hspace{1cm}\beta_k=\frac{\epsilon_k-m}{\sqrt{k^2+\left(\epsilon_k-m\right)^2}}, \label{albe}
\eea
and $\epsilon_k=\sqrt{k^2+m^2}$ with the following normalization
$$|\alpha_k|^2+|\beta_k|^2=1.$$
It is also useful to determine the unentangled IR state in terms of eigenvectors of \eqref{HamiltonianF}
\bea
\left[\alpha_k\,\chi_1(k)+\beta_k\,\chi_2(k)\right]\left| \Omega\right \rangle=\left[-\beta_k\,\chi^{\dagger}_1(k)+\alpha_k\,\chi^{\dagger}_2(k)\right]\left| \Omega\right \rangle=0.
\eea
We want to relate the unentangled IR state to the UV state via the unitary transformation \eqref{unitaryTrnsf}. We will do so by assuming the following disentanglers
\bea
\hat{K}(u)=i\int dk\left[g_k(u)\chi_1^\dagger(k)\chi_2(k)+g^*_k(u)\chi_1(k)\chi_2^\dagger(k)\right]
\eea
where we choose $g_k(u)$, which is generally a complex-value function, to be of the following form
$$g_k(u)=g(u)\Gamma\left(ke^{-u}/\Lambda\right)\frac{k e^{-u}}{\Lambda}.$$
Note that this choice of $g_k(u)$, which is different from the scalar case, is necessary to get the ground state \eqref{Fgrnd} in an approximation justified in the IR region $k<<\Lambda$.
We can now define the creation and annihilation operators at arbitrary energy scale $u$ by the following
\bea
P e^{-i\int_{u_{\mathrm{IR}}}^u\hat{K}(s)ds}\begin{pmatrix}\chi_1(k)\\\chi_2(k)\end{pmatrix}\tilde{P} e^{i\int_{u_{\mathrm{IR}}}^u\hat{K}(s)ds}&=&M_k(u)\begin{pmatrix}\chi_1(k)\\\chi_2(k)\end{pmatrix},\\
P e^{-i\int_{u_{\mathrm{IR}}}^u\hat{K}(s)ds}\begin{pmatrix}\chi^\dagger_1(k)\\\chi^\dagger_2(k)\end{pmatrix}\tilde{P} e^{i\int_{u_{\mathrm{IR}}}^u\hat{K}(s)ds}&=&N_k(u)\begin{pmatrix}\chi^\dagger_1(k)\\\chi^\dagger_2(k)\end{pmatrix}.
\eea
The matrices $M_k(u)$ and $N_k(u)$ have the following form
\bea\label{M&Ndef}
M_k(u)&\equiv&\tilde{P}\exp{\int_{u_{\mathrm{IR}}}^uG_k(s)ds}=\begin{pmatrix}P_k(u)&Q_k(u)\\-Q^*_k(u)&P^*_k(u)\end{pmatrix},\\
N_k(u)&\equiv&\tilde{P}\exp{\int_{u_{\mathrm{IR}}}^uH_k(s)ds}=\begin{pmatrix}P^*_k(u)&Q^*_k(u)\\-Q_k(u)&P_k(u)\end{pmatrix},
\eea
where we are interested in a gauge that $G_k(u)$ and $H_k(u)$ are defined by
\bea\label{G&Hku}
G_k(u)=\begin{pmatrix}0&-g_k(u)\\g^*_k(u)&0\end{pmatrix},\hspace{1cm}H_k(u)=\begin{pmatrix}0&-g^*_k(u)\\g_k(u)&0\end{pmatrix}.
\eea
Note that $M_k(u)$ and $N_k(u)$ both preserve the anti-commutation relations of $\chi$ fields. Also note that we have $|P_k(u)|^2+|Q_k(u)|^2=1$ which together with
$$M_k(u)M^{\dagger}_k(u)=1,\hspace{1cm}N_k(u)N^\dagger_k(u)=1,$$
is showing that these unitary transformations belong to $SU(2)$ and we absorb the effect of the third generator in the phase ambiguity between $A_k$ and $B_k$.
\subsection{Quantum quench}
In a very similar way of what we did for the free scalar field, an excited state due to quantum quench can be approximated by the following boundary state $\left|B\right \rangle$
\ba
&& |\Psi(0)\lb=e^{-\f{\beta}{4}H}|B\lb \no
&& ={\cal{N}}\cdot\exp\left[\pm\int dk_x\,e^{-\beta\epsilon_k/2}\,\chi_1^\dagger(k)\chi_2(k)\right]|0\lb
\ea
where $+$ and $-$ signs correspond to Neumann and Dirichlet boundary conditions. Note that we could have defined the boundary state with $\psi$ fields, instead of $\chi$ fields, where in that case the vacuum state $\left|0\right \rangle$ should be replaced by the IR state $\left|\Omega\right \rangle$ defined above. Using \eqref{quantization} and \eqref{eigenvector} the above UV mixed state can be determined as
\bea
\left[A_k\chi_1(k)+B_k\chi_2(k)\right]|\Psi(0)\lb=0,\hspace{1cm}\left[-B_k\chi^\dagger_1(k)+A_k\chi^\dagger_2(k)\right]|\Psi(0)\lb=0
\eea
where
\bea
A_k=\frac{1}{\sqrt{1+e^{-\beta\epsilon_k}}},\hspace{1cm}B_k=\mp\frac{e^{-\beta\epsilon_k/2}}{\sqrt{1+e^{-\beta\epsilon_k}}},
\eea
and we have normalized by $|A_k|^2+|B_k|^2=1$. In the following we will discuss about the $g(u)$ function after considering the more general case of time dependent quantum quenches.
\subsection{Time dependent excited state}
The time evolution for the free fermion is simply defined similar to the case of free scalar by
\ba
&& |\Psi(0,t)\lb=e^{-\f{\beta}{4}H}|B\lb \no
&& ={\cal{N}}\cdot\exp\left[\pm\int dk_x\,e^{-\beta\epsilon_k/2}e^{-2i\epsilon_k t}\,\chi_1^\dagger(k)\chi_2(k)\right]|0\lb.
\ea
In this case $A_k$ and $B_k$ take the following form
\bea
A_k=\frac{1}{\sqrt{1+e^{-\beta\epsilon_k}}}\,e^{i\epsilon_kt+i\theta_k(t)},\hspace{1cm}B_k=\mp\frac{e^{-\beta\epsilon_k/2}}{\sqrt{1+e^{-\beta\epsilon_k}}}\,e^{-i\epsilon_kt+i\theta_k(t)},
\eea
where $\theta_k(t)$ is again the ambiguity between $A_k$ and $B_k$ which does not change the UV state, but as in the scalar case the intermediate states depend on it.

From now on we will focus on the case of $u=0$ that one can easily find the following from \eqref{M&Ndef}
\bea\label{IntgrEq}
G\left(\log{\frac{k}{\Lambda}}\right)=\int_{\log\frac{k}{\Lambda}}^0\mathcal{G}_k(u)du-k\frac{dM_k(0)}{dk}\cdot M^{-1}_k(0)
\eea
where $\mathcal{G}_k(u)=M_k(u)G_k(u)M^{-1}_k(u)$. Note that the non-Abilean structure makes this equation complicated comparing it to its counterpart in the scalar case.

Since the UV and IR states are related by
\bea
\left(A_k,B_k\right)=\left(\alpha_k,\beta_k\right)\cdot M_k(0),
\eea
$M_k(0)$ could be found as
\bea
P_k(0)=\alpha_kA_k+\beta B^*_k,\hspace{1cm}Q_k(0)=\alpha_kB_k-\beta_kA^*_k.
\eea
Note that our choice (\ref{albe}) does not satisfy the relation $M_\Lambda(0)=1$, which immediately comes from the definition of $M_k(u)$. However, this is not an important problem as long as we focus on the IR region $k<<\Lambda$. This claim was also confirmed in \cite{cMERA} from a different argument. Thus we will focus on
this IR region below, assuming that there is a modification of $|\Omega\lb$ in the UV region so that  $M_\Lambda(0)=1$ is satisfied.\footnote{This is related to the observation that (\ref{albe}) does not include the UV divergences $\Lambda$, while analogous expressions in free scalar field theories do include $\Lambda$ as in  (\ref{irs}).}

We will choose $\theta_k(t)$ such that the form of \eqref{G&Hku} is preserved. The diagonal and off-diagonal components of integral equation \eqref{IntgrEq} correspondingly leads to
\bea
-\frac{\p\theta_k}{\p k}&=&\frac{kt}{\epsilon_k}\tanh\left(\beta\epsilon_k/2\right)+i\frac{\epsilon_k}{mk}\int_{\log\frac{|k|}{\Lambda}}^0\left[\mathcal{G}_k(u)\right]_{11}du\pm \mathcal{H}_1(k,t)\label{Eqtheta},\\
g\left(\log{\frac{k}{\Lambda}}\right)&=&\frac{1}{2}\frac{mk}{\epsilon_k^2}-\int_{\log\frac{|k|}{\Lambda}}^0\left\{\left[\mathcal{G}_k(u)\right]_{12}+\frac{k}{m}\left[\mathcal{G}_k(u)\right]_{11}\right\}du\pm \mathcal{H}_2\left(k,t\right),
\label{Eqg}
\eea
where
\bea
\mathcal{H}_1(k,t)&=&\frac{k^2}{4 m\epsilon_k \cosh\left(\beta\epsilon_k/2\right)}\left[4 t \cos2\theta_k+\beta\sin2\theta_k\right],\\
\mathcal{H}_2(k,t)&=&\frac{k^2}{4\epsilon_k\cosh\left(\beta\epsilon_k/2\right)}\left(\beta\cos2\theta_k-4t\sin2\theta_k\right)+i\epsilon_k\mathcal{H}_1(k,t).
\eea
This complicated equations arise because of the non-abelian group structure which is manifest in \eqref{G&Hku}. To study the time dependence of $g(u)$, one has to solve these integral equations \eqref{Eqtheta} and \eqref{Eqg} at least numerically. In what follows we will just focus on a simple case of $t=0$, which corresponds to just after the quantum quench. In this case $g(u)$ is a real valued function and we can forget about the non-abelian structure. However, in the massless theory, it is clear from the above expressions that $g(u)$ scales like $t/\beta$ at late time for $k\beta<<1$. Thus the behavior of entanglement entropy and the holographic metric are qualitatively similar to those for the free scalar theories discussed previously.

\subsection{Metric just after quantum quench}
\begin{figure}[ttt]
   \begin{center}
     \includegraphics[scale=.75]{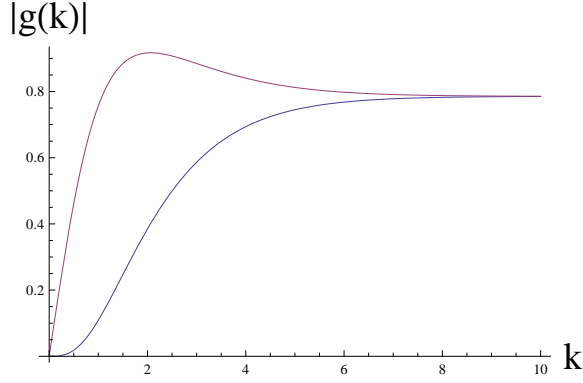}
   \end{center}
   \caption{The plot of $|g(u)|$ of massless free fermions ($m=0$) just after the quench for Neumann (blue curve) and Dirichlet (red curve) boundary conditions. We have chosen $\beta=2$.}\label{fig:gF}
\end{figure}

As mentioned above, in this case $g(u)$ is a real function, thus we can replace $\mathcal{G}_k(u)$ by $G_k(u)$ in \eqref{Eqtheta} and \eqref{Eqg} and we can easily find
\bea
M_k(u)=\begin{pmatrix}\cos\phi_k(u)&-\sin\phi_k(u)\\\sin\phi_k(u)&\cos\phi_k(u)\end{pmatrix}
\eea
where
\bea
\phi_k(u)=\int_{u_{\mathrm{IR}}}^ug_k(s)ds.
\eea
Thus in the case of $u=0$ one can find
\bea
\sin2\phi_k=-\frac{k\tanh\left(\beta\epsilon_k/2\right)\mp m\,\mathrm{sech}\left(\beta\epsilon_k/2\right)}{\epsilon_k}
\eea
which leads to
\bea
g\left(u\right)&=&\frac{1}{2}\frac{m\Lambda e^u}{\epsilon_u^2}-\frac{1}{2}\arcsin\left[\frac{\Lambda e^u\tanh\left(\beta\epsilon_u/2\right)\pm m\,\mathrm{sech}\left(\beta\epsilon_u/2\right)}{\epsilon_u}\right]\nonumber\\&\pm&\frac{\Lambda^2 e^{2u}\beta}{4\epsilon_u\cosh\left(\beta\epsilon_u/2\right)}, \label{metfqm}
\eea
with the identification $k=\Lambda e^u$. Note that the above $g(u)$ in the $\beta\to\infty$ limit reduces to the known result previously discussed in \cite{NRT}. This function is plotted for both boundary conditions in Fig.\ref{fig:gF}. Following the arguments in \cite{NRT}, again we find the holographic metric as $g_{uu}(u)\propto g(u)^2$.

\section{Finite Chemical Potential}

 Finally we would like to discuss a generalization of our finite temperature cMERA by including a chemical potential $\mu$. We will study both free scalar and fermion theory. Especially, in the latter theory, we will find a sharp peak of
$g(u)$ at the fermi level.

\subsection{Free Scalar Field Theory with Chemical Potential}

For this purpose, in the free scalar field example, we need to replace (\ref{thwave}) with\footnote{One may think that we should consider a complex scalar field in order to have a charged field. However, this is equivalent to
a real scalar with the chemical potential $\mu$ and that with $-\mu$. In this sense we can directly apply our argument below to this complex scalar field theory.}
\ba
&& |\Psi(0,t)\lb_{th}={\cal N}\cdot e^{-it(H_1+H_2)}\cdot \prod_{k}\sum_{n_k=0}^\infty e^{-\beta (\ep_k-\mu) n_k/2}
|n_k\lb_1 |n_k\lb_2 \no
&& \ \ ={\cal N}\cdot \exp\left(\int dk e^{-\f{\beta (\ep_k-\mu)}{2}}e^{-2i\ep_k t}a^{\dagger}_k
\ti{a}^{\dagger}_k  \right)|0\lb |\ti{0}\lb.  \label{thh}
\ea
We can introduce the parameter $\mu$ for quantum quench in the same way.

\begin{figure}[ttt]
   \begin{center}
     \includegraphics[height=4cm]{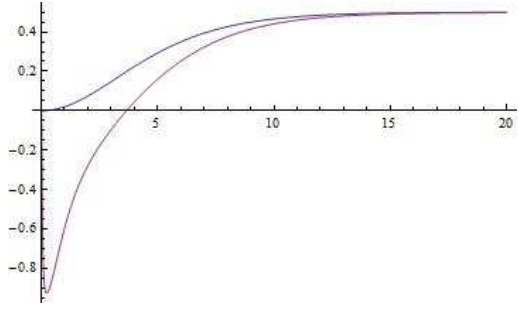}
   \end{center}
   \caption{The plot of $g(u)$ (red curve) for the massive scalar field as a function of $k=\Lambda e^u$. We chose $m=\beta=1$ and $\mu=0.999$. We also inserted the plot of $g(u)$ (blue curve) for  the massless scalar field for $\beta=1$ as a reference.}\label{fig:mu}
\end{figure}

Since we encounter the divergence when $\ep_k<\mu$, we need to assume a non-vanishing scalar field mass $m$ and restrict the values of chemical potential in the range $|\mu|<m$.
 Let us focus on the $t=0$ state. The function $g(u)$ is determined as follows
\be
g(u)=\f{k^2}{2(k^2+m^2)}-\f{\beta k^2}{4\s{k^2+m^2}\sinh\f{\beta(\s{k^2+m^2}-\mu)}{2}}.
\ee
It is interesting to note that the absolute value of this function approaches $|g(u)|=1$
when $k$ is very small if $\mu$ gets very close to $m$ as we showed in Fig.\ref{fig:mu}.
Remembering that the metric $g_{uu}$ is proportional to $g(u)^2$, this suggests that in the charged thermal system, $g_{uu}$ gets larger in the IR region.

\subsection{Free Fermion Theory with Chemical Potential}

Now we turn to the free Dirac fermion theory with the chemical potential $\mu$. In this case, there is no constraint for the values of $\mu$. For simplicity we assume the massless limit $m=0$ and then the function $g(u)$, which is proportional to $\s{g_{uu}}$, is evaluated by slightly modifying (\ref{metfqm}) as follows:
\be
g\left(u\right)=-\frac{1}{2}\arcsin\left[\tanh\left(\beta (\Lambda e^u-\mu) /2\right)\right]\pm \frac{\Lambda e^{u}\beta}{4\cosh\left(\beta(\Lambda e^u-\mu)/2\right)}.
\ee

\begin{figure}[ttt]
   \begin{center}
     \includegraphics[height=4cm]{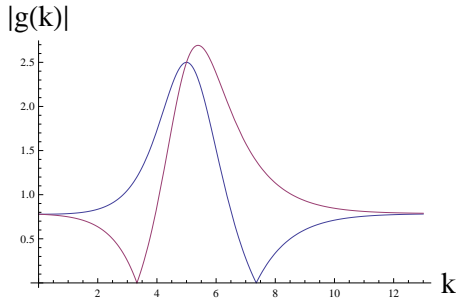}
   \end{center}
   \caption{The plots of $|g(u)|$ for the massless fermion field as a function of $k=\Lambda e^u$, for the two boundary conditions. We chose $\beta=2$ and $\mu=5$.}\label{fig:Fmu}
\end{figure}

In the UV limit $u\to 0$, we get the same result $g(0)=-\f{\pi}{4}$ as that for $\mu=0$.
On the other hand, in the IR region we find a non-trivial effect of finite $\mu$. In general we find that $|g(u)|$ enhances in the IR (see Fig.\ref{fig:Fmu}). Especially there is a peak at
$\Lambda e^u(=k)=\mu$ i.e. at the fermi level and there $g(u)$ is estimated as
$|g(u)|\simeq \f{\beta\mu}{4}$, which can be taken to be arbitrary large by choosing a large value of $\mu$. This seems to be analogous to a changed extremal black hole in AdS spaces, where the fermi surface scale $z=1/\mu$ can be related to the black hole horizon as expected from the AdS/CFT.

\section{Conclusions}

In this paper, we studied constructions and properties of cMERA beyond ground states by focusing on the free field theories. We analyzed cMERA for excited states defined by quantum quenches and computed its holographic metric $g_{uu}$ in the extra dimension. From the view point of cMERA, this metric measures how much quantum entanglement exists at a given length scale.
The time evolution of this metric looks like gravitational wave propagations inside an analogue of light cone. We found that there is a non-trivial region where the metric gets very large and we identified it as the region inside the horizon. Indeed, the cMERA result shows that this region grows linearly with the time evolution, which is consistent with the proposed gravity dual of quantum quenches.

Moreover, motivated by the relation between (both holographic and field theoretic) descriptions of quantum quenches and finite temperature CFTs, we proposed a cMERA construction at finite temperature. The holographic metric calculated in this cMERA construction agrees with the gravity dual prediction.

Finally we analyzed the cMERA in the presence of chemical potential. We found a new behavior that the metric gets large in the IR region. Especially in fermion theories, we find that the metric can have a large peak at the fermi surface, which might be related to the extremal black hole horizon.

There are many future problems. Since we have considered free field theories, we encountered the oscillations of metric function $g_{uu}$. We believe this is an artifact of free field theory because in interacting theories, the sectors with different momentum are mixed by the interactions and the oscillations with a fixed value of momentum will be washed out. It will be an important future problem to confirm this explicitly. It will also be interesting to find a cMERA description of localized excitations instead of the translationally invariant ones discussed in this paper.
This will be related to the local quenches \cite{CaCAL} and there have been examples of their gravity duals \cite{localQ}. One more intriguing future problem is to explore universal properties of excited
states from the viewpoint of cMERA
(refer to recent results from holography and CFT calculations \cite{excited}).

\section*{Acknowledgements}

We would like to thank Jyotirmoy Bhattacharya, Veronika Hubeny, Rene Meyer, Robert Myers, Hirosi Ooguri, Mukund Rangamani, Simon Ross and Tomonori Ugajin for useful discussions and comments. AM would also like to thank Mohsen Alishahiha and Mohammad Reza Mohammadi-Mozaffar for discussions about basic topics related to this work. AM would like to thank Yukawa Institute for Theoretical Physics (YITP) for hospitality during this work. MN is thankful very much to the Perimeter Institute for
kind hospitality where a part of this work was done.
TT is very grateful to the scientific program ``Mathematics and Physics of the Holographic Principle,'' at Isaac Newton Institute for Mathematical Sciences and to the Centre for Particle Theory
at Durham University for wonderful hospitality, where this work was partly done.
SR has been supported
by the U.S. Department of Energy, Office of Basic Energy Sciences, Division of Material Sciences and Engineering (DE-FG02-12ER46875).
TT is supported by JSPS Grant-in-Aid for Scientific
Research (B) No.25287058 and JSPS Grant-in-Aid for Challenging
Exploratory Research No.24654057. AM is supported by Iran Ministry of Science, Research and Technology grant for PhD students sabbatical. TT is also
supported by World Premier International
Research Center Initiative (WPI Initiative) from the Japan Ministry
of Education, Culture, Sports, Science and Technology (MEXT).

\appendix

\section{Another Definition of Metric for
Finite Temperature cMERA}

The UV density matrix for the finite temperature free scalar field theory in $d+1$ dimension is given by
\begin{equation}
\begin{split}
\left|\Psi_{UV} \right\rangle &=\frac{1}{Z(\beta)^{\frac{1}{2}}}\exp{\left( \int d^dk e^{-\frac{\epsilon_k \beta}{2}}a_k^{\dagger}\tilde{a}_k^{\dagger}\right)}\left| 0,\tilde{0} \right\rangle, \\
Z(\beta) &= Tr e^{-\beta \mathcal{H}}.
\end{split}
\end{equation}
The state at energy scale $u$ (Used math mode for u) is given by
\begin{equation}
\begin{split}
\left|\Psi(u)\right \rangle &= e^{-i u L }\mathcal{\tilde{P}}e^{\left( i \int^{0}_{u}\hat{K}_1(s)ds\right)}\cdot \mathcal{\tilde{P}} e^{\left( i \int^{0}_{u}\hat{K}_1(s)ds\right)}\left|\Psi_{UV} \right \rangle, \\
K_{1}(s)&=\frac{i}{2}\int d^dk\int_{\left| k \right|\le \Lambda e^s}g(s)\left(a_k^{\dagger}a_{-k}^{\dagger}-a_k a_{-k}\right), \\
K_{2}(s)&=\frac{i}{2}\int d^dk\int_{\left| k \right|\le \Lambda e^s}g(s)\left(\tilde{a}_k^{\dagger}\tilde{a}_{-k}^{\dagger}-\tilde{a}_k \tilde{a}_{-k}\right).
\end{split}
\end{equation}
$\left| \Phi(u) \right\rangle$ is given by
\begin{equation}
\left| \Phi(u) \right\rangle = e^{i u L}\left|\Psi(u)\right\rangle.
\end{equation}
The metric along the extra holographic direction is given by
\begin{equation}
\begin{split}
g_{uu}(u)&=\mathcal{N}^{-1}\left( 1-\left| \left\langle \Psi(u) \right| e^{i du \cdot L}\left| \Psi(u+du)\right\rangle \right|^2 \right)=\mathcal{N}^{-1}\left( 1-\left| \left\langle \Phi(u) \big{|} \Phi(u+du)\right\rangle \right|^2 \right) \\
&\sim \big{[}2\left\langle \Phi(u)\right|K_1(u)K_{2}(u)\left|\Phi(u)\right\rangle +\left\langle \Phi(u)\right|\left(K_1^2(u)+K_2^2(u)\right)\left|\Phi(u)\right\rangle \\
&~~~~~~~~~~~~~~~~~~~~~~~~~~~~~~~~~~~~~~~~~~~~~~~~~~~~- \left|\left\langle \Phi(u)\right|\left(K_1(u)+K_2(u)\right)\left|\Phi(u)\right\rangle\right|^2\big{]}du^2 \\
&=\left[ 2\left\langle\Psi_{UV}\right|K_1(u)K_2(u)\left|\Psi_{UV}\right\rangle +\left\langle\Psi_{UV}\right|K_1^2(u)+K_2^2(u)\left|\Psi_{UV}\right\rangle\right]du^2
\end{split}
\end{equation}
where
\begin{equation}
\left\langle \Phi(u)\right|\left(K_1(u)+K_2(u)\right)\left|\Phi(u)\right\rangle=\left\langle \Psi_{UV}\right|\left(K_1(u)+K_2(u)\right)\left|\Psi_{UV}\right\rangle=0.
\end{equation}

$K_1^2(u)$ is given by
\begin{equation}
\begin{split}
K_1^2(u)&=\frac{-1}{4}\int_{\left|k\right|\le\Lambda\cdot e^u}d^dk\int_{\left|p\right|\le\Lambda\cdot e^u}d^dp~g^2(u) \\
&\times \left(a_k^{\dagger}a_{-k}^{\dagger}a_p^{\dagger}a_{-p}^{\dagger}-a_k^{\dagger}a_{-k}^{\dagger}a_p a_{-p}-a_k a_{-k} a_p^{\dagger}a_{-p}^{\dagger}+a_k a_{-k} a_p a_{-p} \right). \\
\end{split}
\end{equation}
\begin{equation}
\begin{split}
\left\langle\Psi_{UV}\right|K_1^2(u)\left|\Psi_{UV}\right\rangle &= \frac{1}{4}\int_{\left|k\right|\le\Lambda\cdot e^u}d^dk\int_{\left|p\right|\le\Lambda\cdot e^u}d^dp ~g^2(u) \\
& \times \left\langle\Psi_{UV}\right|\left( a_k^{\dagger}a_{-k}^{\dagger}a_p a_{-p} +a_k a_{-k} a_p^{\dagger} a_{-p}^{\dagger} \right)\left|\Psi_{UV}\right\rangle \\
&=\frac{g^2(u)}{4 Z(\beta)}\int_{\left|k\right|\le\Lambda\cdot e^u}d^dk\int_{\left|p\right|\le\Lambda\cdot e^u}d^dp tr\left(\rho~ a^{\dagger}_k a^{\dagger}_{-k}a_{p}a_{-p}+\rho~ a_{k}a_{-k}a^{\dagger}_{p}a^{\dagger}_{-p}\right), \\
\end{split}
\end{equation}
where
\begin{equation}
tr\left(\rho~a^{\dagger}_{k}a^{\dagger}_{-k}a_{p}a_{-p}\right)= tr\left(a_p a_{-p}~\rho~a^{\dagger}_k a^{\dagger}_{-k}\right)= tr\left(\rho a_p a_{-p}a^{\dagger}_k a^{\dagger}_{-k}\right)e^{-\beta \epsilon_p}.
\end{equation}
Then,
\begin{equation}
\begin{split}
\left\langle\Psi_{UV}\right|K_1^2(u)\left|\Psi_{UV}\right\rangle &=\frac{g^2(u)}{4 Z(\beta)}\int_{\left|k\right|\le\Lambda\cdot e^u} d^dk\int_{\left|p\right|\le\Lambda\cdot e^u} d^dp \left(1+e^{-2\beta \epsilon_k} \right)tr\left(e^{-\beta \mathcal{H}}a_k a_{-k}a^{\dagger}_p a^{\dagger}_{-p}\right)  \\
&=\frac{g ^2(u)}{4}\int_{\left|k\right|\le\Lambda\cdot e^u} d^dk \int_{\left|p\right|\le\Lambda\cdot e^u} d^dp \left(1+e^{-2\beta \epsilon_k}\right) \\
&~~~~~~~~~~~~~~~~~~~~\times \frac{\left(\big{[}a_{-k},a^{\dagger}_{p}\big{]} \cdot \left[a_{k},a^{\dagger}_{-p}\right]+\left[a_{k},a^{\dagger}_{p}\right]\cdot\left[a_{-k},a^{\dagger}_{-p}\right]\right)}{\left(1-e^{-\beta \epsilon_k}\right)^2} \\
&=\frac{g^2(u)}{4}\int_{\left|k\right|\le\Lambda\cdot e^u} d^dk\frac{2\left( 1+e^{-2\beta} \right)}{\left(1-e^{-\beta \epsilon_k}\right)^2}.
\end{split}
\end{equation}
Then,
\begin{equation}
\left\langle\Psi_{UV}\right|\left( K_1^2(u)+K_2^2(u)\right)\left|\Psi_{UV}\right\rangle=g^2(u) \int_{\left|k\right|\le\Lambda\cdot e^u} d^dk\frac{\delta (0)\left( 1+e^{-2\beta} \right)}{\left(1-e^{-\beta \epsilon_k}\right)^2}.
\end{equation}

\begin{equation}
\begin{split}
K_1(u)\cdot K_2(u)&=\frac{-g^2 (u)}{4}\int_{\left|p\right|\le\Lambda\cdot e^u}d^dp\int_{\left|k\right|\le\Lambda\cdot e^u}d^dk \\
&\times \left\{ a^{\dagger}_{p}a^\dagger_{-p}\tilde{a}^{\dagger}_{k}\tilde{a}^\dagger_{-k}-a^{\dagger}_{p}a^\dagger_{-p}\tilde{a}_{k}\tilde{a}_{-k}-a_{p}a_{-p}\tilde{a}^{\dagger}_{k}\tilde{a}^\dagger_{-k}+a_{p}a_{-p}\tilde{a}_{k}\tilde{a}_{-k}\right\}.
\end{split}
\end{equation}
The second term and third term disappear.
\begin{equation}
\begin{split}
\left\langle \Psi_{UV}\right| & \left( a^{\dagger}_p a^{\dagger}_{-p} \tilde{a}^{\dagger}_k \tilde{a}^{\dagger}_{-k}+a_p a_{-p}\tilde{a}_k \tilde{a}_{-k}\right)\left| \Psi_{UV} \right\rangle =\frac{tr\left\{e^{-\beta \mathcal{H}}\left( a^{\dagger}_p a^{\dagger}_{-p} \tilde{a}^{\dagger}_k \tilde{a}^{\dagger}_{-k}+a_p a_{-p}\tilde{a}_k \tilde{a}_{-k}\right)\right\}}{Z(\beta)} \\
&=\frac{tr\left( e^{-\beta \mathcal{H}}a_k a_{-k}a^{\dagger}_p a^{\dagger}_{-p}\right)e^{-\beta \epsilon_k}+ tr\left( e^{-\beta \mathcal{H}}a_p a_{-p}a^{\dagger}_k a^{\dagger}_{-k}\right)e^{-\beta \epsilon_k}}{Z(\beta)} \\
&=2\left \{ \frac{\left(\big{[}a_{-k},a^{\dagger}_{p}\big{]} \cdot \left[a_{k},a^{\dagger}_{-p}\right]+\left[a_{k},a^{\dagger}_{p}\right]\cdot\left[a_{-k},a^{\dagger}_{-p}\right]\right)}{\left(1-e^{-\beta \epsilon_k}\right)^2} \right\}e^{-\beta \epsilon_k}.
\end{split}
\end{equation}
Then, we find
\begin{equation}
\begin{split}
&\left\langle\Psi_{UV}\right|K_1(u)K_2(u)\left|\Psi_{UV}\right\rangle
= -g^2(u)\int_{\left|p\right|\le\Lambda\cdot e^u} d^dp\frac{\delta(0)e^{-\beta \epsilon_k}}{(1-e^{-\beta \epsilon_k})^2} \\
\end{split}
\end{equation}

\begin{equation}
\begin{split}
&2\left\langle\Psi_{UV}\right|K_1(u)K_2(u)\left|\Psi_{UV}\right\rangle + \left\langle\Psi_{UV}\right|K_1^2(u)+K_2^2(u)\left|\Psi_{UV}\right\rangle =g^2(u)\cdot \mathcal{N}, \\
&\mathcal{N}=\int_{\left|p\right|\le\Lambda\cdot e^u} d^dp
\end{split}
\end{equation}

Finally we obtain
\begin{equation}
g_{uu}(u)=\chi^2(u)
\end{equation}

\end{document}